\documentclass[final,5p,times,twocolumn,sort&compress]{elsarticle}
\usepackage{float}
\usepackage{lineno}

\usepackage{graphicx}
\usepackage{amssymb,amsmath,bm}
\usepackage{amsthm}
\usepackage{txfonts}
\usepackage{mathtools, cuted}
\usepackage{lipsum, color}
\usepackage{subcaption}
\usepackage{color}
\usepackage[utf8]{inputenc}
\usepackage{amsmath}
\usepackage{graphicx}
\usepackage{wrapfig}
\usepackage{setspace}
\usepackage{listings,newtxtt}
\usepackage{dirtree}
\usepackage{makecell}
\usepackage{hyperref}
\usepackage{multirow}
\usepackage{amsfonts}
\usepackage{color}
\usepackage{cuted}
\usepackage{svg}
\usepackage{tikz}
\PassOptionsToPackage{hyphens}{url}\usepackage{hyperref}

\definecolor{L25}{rgb}{0,0.25,1}
\definecolor{L50}{rgb}{0.5,1,0.5}
\definecolor{L100}{rgb}{1,0.75,0}
\definecolor{L200}{rgb}{1,0,0}
\definecolor{HPAM2}{rgb}{0.168627451,0.7568627451,0.6078431373}
\definecolor{HPAM1}{rgb}{0.7529411765,0.8705882353,0.1960784314}

\newrobustcmd*{\mysquare}[1]{\tikz{\filldraw[draw=#1,fill=#1] (0,0)
rectangle (0.2cm,0.2cm);}}

\newrobustcmd*{\mysquareExp}[1]{\tikz{\filldraw[draw=black,fill=#1] (0,0)
rectangle (0.2cm,0.2cm);}}

\newrobustcmd*{\mycircle}[1]{\tikz{\filldraw[draw=#1,fill=#1] (0,0) circle [radius=0.1cm];}}

\newrobustcmd*{\mycircleExp}[1]{\tikz{\filldraw[draw=black,fill=#1] (0,0) circle [radius=0.1cm];}}

\newrobustcmd*{\mytriangleExp}[1]{\tikz{\filldraw[draw=black,fill=#1] (0,0) --
(0.2cm,0cm) -- (0.1cm,0.1732050808cm);}}

\newrobustcmd*{\mytriangle}[1]{\tikz{\filldraw[draw=#1,fill=#1] (0,0) --
(0.2cm,0cm) -- (0.1cm,0.1732050808cm);}}

\journal{Journal of Non-Newtonian Fluid Mechanics}

\begin{document}

\begin{frontmatter}


\title{Master Curves for FENE-P Fluids in Steady Shear Flow}



\author[MIT]{Sami Yamani*}
\ead{syamani@mit.edu}
\author[MIT]{Gareth H. McKinley*}
\ead{gareth@mit.edu}
\cortext[corrauth]{Author to whom correspondence should be addressed.}

\address[MIT]{Hatsopoulos Microfluids Laboratory, Department of Mechanical Engineering, Massachusetts Institute of Technology, Cambridge, MA 02139, USA}


\begin{abstract}

The FENE-P (Finitely-Extensible Nonlinear Elastic) dumbbell constitutive equation is widely used in simulations and stability analyses of free and wall-bounded viscoelastic shear flows due to its relative simplicity and accuracy in predicting macroscopic properties of dilute polymer solutions.  The model contains three independent material parameters, which expressed in dimensionless form correspond to a Weissenberg number ($\textrm{Wi}$), \textit{i.e.}, the ratio of the dumbbell relaxation time scale to a characteristic flow time scale, a finite extensibility parameter ($L$), corresponding to the ratio of the fully extended dumbbell length to the root mean square end-to-end separation of the polymer chain under equilibrium conditions, and a solvent viscosity ratio, commonly denoted $\beta$. An exact solution for the rheological predictions of the FENE-P model in steady simple shear flow is available [Sureshkumar \textit{et al.}, Phys Fluids (1997)], but the resulting nonlinear and nested set of equations do not readily reveal the key shear-thinning physics that dominates at high $\textrm{Wi}$ as a result of the finite extensibility of the polymer chain. In this note we review a simple way of evaluating the steady material functions characterizing the nonlinear evolution of the polymeric contributions to the shear stress and first normal stress difference as the shear rate increases, provide asymptotic expansions as a function of $\textrm{Wi}$ , and show that it is in fact possible to construct universal master curves for these two material functions as well as the corresponding stress ratio. Steady shear flow experiments on three highly elastic dilute polymer solutions of different finite extensibilities also follow the identified master curves. The governing dimensionless parameter for these master curves is $\textrm{Wi}/L$ and it is only in strong shear flows exceeding $\textrm{Wi}/L \gtrsim 1$ that the effects of finite extensibility of the polymer chains dominate the evolution of polymeric stresses in the flow field. We suggest that reporting the magnitude of $\textrm{Wi}/L$ when performing stability analyses or simulating shear-dominated flows with the FENE-P model will help clarify finite extensibility effects. 

\end{abstract}



\end{frontmatter}


\section{Introduction}
\label{sec:intro}

Dilute solutions of flexible polymer chains are one of the most common subclasses of non-Newtonian fluids, with a principal characteristic known to be the pronounced nonlinear viscoelasticity generated at high deformation rates. The polymer chains at equilibrium relax to a random coil configuration. These polymer coils, however, can be unravelled and attain stretched configurations based on the strength and type of an imposed flow. Two canonical flow types have been studied extensively for dilute polymer solutions: steady extensional flow and steady shear flow. Here, we focus on steady shear flow. 


The simplest way to model the deformation of a dilute suspension of polymer chains in a solution is to treat each isolated chain as a non-interacting dumbbell composed of a pair of beads that are connected to each other by an entropic spring \cite{bird1987dynamicsv2}. The Hookean spring exerts a restoring force to prevent separation of the two beads in the dumbbell, and a relaxation time scale can be defined based on the spring constant and the friction coefficient of the beads. A force balance based on the viscous drag and linear spring forces characterizes the evolution of the dumbbell connector vector, \textit{i.e.}, the vector connecting the two beads \cite{rallison1988we}. Ensemble averaging of the conformation of each dumbbell leads to an evolution equation for the second moment of the dumbbell connector vector (\textit{i.e.}, the dumbbell conformation tensor) in terms of the upper convected derivative. The resulting closed-form constitutive equation for the total stress (including both polymer and solvent contributions) is identical to the constitutive equation suggested by Oldroyd \cite{oldroyd1950formulation} from continuum mechanical considerations that is commonly referred to as the Oldroyd-B \cite{rallison1988we, hinch2021oldroyd} model. This Hookean dumbbell model was later modified to recognize the finite extensibility of a polymer chain by making the spring connecting the two beads of the dumbbell finitely extensible and nonlinearly elastic (FENE) \cite{reinhold1966hydrodynamics,peterlin1961einfluss,peterlin1968,tanner1971stresses}. 
A closed form constitutive equation for the ensemble average of the second moment tensor FENE model was developed by Peterlin (FENE-P) \cite{bird1987dynamicsv1} using a pre-averaging assumption. The resulting FENE-P constitutive equation, in combination with the continuity equation and Cauchy momentum equation provide a set of non-linearly coupled equations that can be solved to find the velocity and and stress fields in steady or time-dependent flows. This feature makes the FENE-P model a popular choice for continuum mechanics simulations of dilute polymer solutions in complex mixed flows \cite{buza2022weakly,shekar2021tollmien, lopez2019dynamics, guimaraes2020direct, parvar2022steady, lopez2022vortex, guimaraes2022turbulent, dubief2022first}. When reporting the results of simulations with the FENE-P model, three dimensionless numbers are important; (i) the Weissenberg number (Wi), the ratio of the relaxation time scale of the dumbbell to the advection time scale of the flow, (ii) the finite extensibility ($L$), the ratio of the dumbbell length at its fully extended state to the root mean squared polymer size at equilibrium, and (iii) the viscosity ratio ($\beta$), the ratio of the solvent viscosity contribution to the total zero shear rate viscosity of the solution.


The asymptotic response of the FENE dumbbell model can be readily evaluated in an irrotational extensional flow (see for example chapter 13.5 of \cite{bird1987dynamicsv2}). The polymer chains orient with the elongational axis and approach full stretch at high extension rates, in good agreement with experimental observations and numerical simulations \cite{larson2005rheology}. The picture is more complex in strong shear flows due to the presence of vorticity. Bead-rod \cite{doyle1997dynamic, hur2001dynamics} and bead-spring \cite{hsieh2004modeling,lyulin1999brownian, schroeder2005dynamics, schroeder2005characteristic, jendrejack2002stochastic} simulations show that individual chains partially align and stretch in the flow direction but also tumble continuously due to the presence of vorticity. Tumbling was first predicted by Lumley \cite{lumley1969drag} and de Gennes \cite{de1974coil} and later directly observed in experiments on deoxyribonucleic acid (DNA) macromolecules \cite{schroeder2005characteristic, teixeira2005shear}. Experiments and Brownian dynamics simulations show that a partially elongated polymer chain has a non-zero mean orientation angle with respect to the streamwise direction. In this state, the shear flow stretches the polymer and progressively aligns it with the flow (reducing the orientation angle). If the flow is not strong enough (low Weissenberg number), the polymer chain's entropic restoring force overcomes the shearing force and the polymer chain starts to recoil. In strong flows (high Weissenberg number), the mean orientation angle scales as $1/\textrm{Wi}$ and approaches zero as the polymer chains becomes more and more aligned with the flow. However, Brownian fluctuations of parts of the chain can make the local angle with the flow momentarily negative, which is enough to cause the polymer chain  to tumble \cite{schroeder2005characteristic}. As the Weissenberg number increases from small to moderate values, the probability of recoil motion decreases and instead the probability of tumbling motion increases. 

Repeated stretching, tumbling, and recoiling of polymer chains in a strong shear flow give rise to shear thinning and shear-rate-dependent normal stresses in the solution \cite{bird1987dynamicsv1}. These macroscopic rheological features can be modeled by both closed form constitutive equation dumbbell models such as the FENE-P model and also evaluated by Brownian dynamics models of freely jointed bead-rod \cite{doyle1997dynamic, hur2001dynamics} and bead-spring \cite{hsieh2004modeling,lyulin1999brownian, schroeder2005dynamics, schroeder2005characteristic, jendrejack2002stochastic} chains. They can be most clearly quantified by calculating the rate-dependence of the polymer contribution to the solution viscosity ($\eta_{p}$) and the first normal stress coefficient ($\Psi_1$). Simple fluid theory and bead-spring simulations show that both $\eta_p$ and $\Psi_1$ are independent of Weissenberg number for small Weissenberg numbers ($\textrm{Wi} \lesssim 1$), however, they both experience a power-law decay as Weissenberg number increases to moderate and high values. For moderate values of Weissenberg number ($ 10 \lesssim \textrm{Wi} \lesssim 100$), the bead-spring model predicts power-laws of $-1/2$ and $-14/11$ for the polymer viscosity and first normal stress coefficient, respectively \cite{doyle1997dynamic,lyulin1999brownian,hsieh2004modeling,jendrejack2002stochastic,schroeder2005characteristic}. At high values of Weissenberg number ($\textrm{Wi} \gtrsim 100$), however, the numerical simulations approach the classical FENE dumbbell result \cite{schroeder2005dynamics} and the power-laws change to $-2/3$ and $-4/3$ for the polymer viscosity and first normal stress coefficient, respectively \cite{doyle1997dynamic,lyulin1999brownian,hsieh2004modeling,jendrejack2002stochastic,schroeder2005characteristic}. It is generally observed that at high Weissenberg numbers, the rheological response of the solution is dominated by the finite extensibility of the springs and other parameters such as hydrodynamic interactions between beads and excluded volume do not play a significant role \cite{larson2005rheology}.
 
In this work, we examine the predictions of the FENE-P constitutive equation for finitely extensible dumbbells in steady shear flow. Using asymptotic analysis for low and high limits of the Weissenberg number, we show that there is a previously unreported master curve that collapses the model predictions for material functions such as the polymer viscosity and first normal stress coefficient for all values of the Weissenberg number and independent of the finite extensibility parameter. Our analysis shows that $\textrm{Wi}/L$ is the governing dimensionless number that characterizes the flow conditions when the rheological properties of a FENE-P fluid are dominated by finite extensibility. Steady shear experiments with dilute polymer solutions in viscous solvents are then carried out showing results that are in reasonable agreement with the FENE-P master curves even though the FENE-P dumbbell model is a greatly simplified model of real flexible polymer chains. 

\section{Governing equations}
\label{sec:GovEq}
\begin{figure}[tbhp]
    \centering
    \includegraphics[width=0.5\columnwidth]{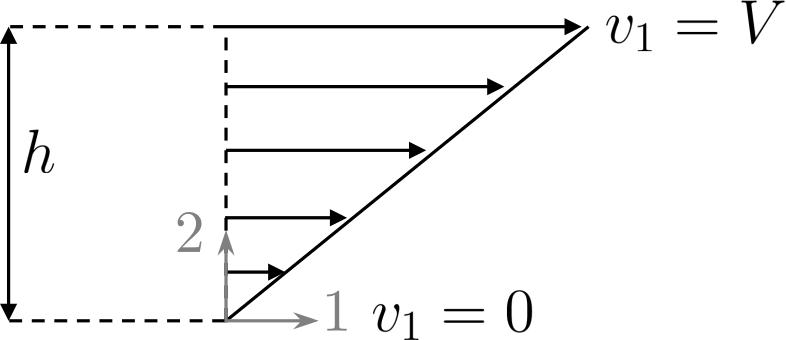}
    \caption{Velocity profile in steady simple shear flow.}
    \label{fig:Shear}
\end{figure}

In this study, we focus on understating the behavior of the FENE-P constitutive model \cite{bird1987dynamicsv1} in steady simple shear flow, shown schematically in Fig.\,\ref{fig:Shear}. The streamwise velocity is $v_1 = 0$ at spanwise position $x_2 = 0$ and linearly grows to $v_1 = V$ at spanwise position $x_2 = h$. A homogeneous  state shear rate can thus be defined as $\dot{\gamma} = V/h$.

The governing equations of the FENE-P fluid in steady shear flow are the continuity equation for an incompressible fluid and the Cauchy momentum equation. The stress tensor in the Cauchy momentum equation is decomposed into the stress contribution from the solvent, assumed to be a Newtonian incompressible fluid, and the stress contribution from the polymer molecules that are modeled by the FENE-P constitutive model. The Cauchy momentum equation can be written as, 



\begin{equation}
 \rho \frac{D}{Dt}\mathbf{v} = -\boldsymbol{\nabla}P + \beta \eta_0 \mathbf{\nabla}^2 \mathbf{v} + \boldsymbol{\nabla} \cdot \boldsymbol{\tau}, 
  \label{eq:Cauchy}
\end{equation}
where $D \mathbf{v} /Dt = \partial \mathbf{v}/\partial t + \mathbf{v} \cdot \boldsymbol{\nabla} \mathbf{v}$ is the material derivative of the velocity vector, $\mathbf{v}$ \cite{bird1987dynamicsv1}. In Eq.\,(\ref{eq:Cauchy}), $P$ is the pressure, $\rho$ is the fluid density, $\eta_0$ is the zero shear rate viscosity of the fluid, $\beta = \eta_s/\eta_0$ is the ratio of the solvent viscosity to the zero shear rate viscosity of the solution, commonly termed the viscosity ratio, and $\eta_s$ is the solvent viscosity such that $\eta_0 = \eta_s + \eta_p(0)$, where $\eta_p(0)$ is the polymer viscosity in the limit of zero shear rate, $\dot{\gamma} \rightarrow 0$, in the FENE-P model. The polymer stress contribution to the Cauchy momentum equation ($\boldsymbol{\tau}$) is a function of the polymer chain conformation tensor $\mathbf{A} = \langle \mathbf{r} \mathbf{r}\rangle/\left(\langle r_0^2 \rangle/3\right)$, where $\mathbf{r}$ is the dumbbell connector vector and $\langle r_0^2 \rangle$ is the mean squared end-to-end distance of a dumbbell at equilibrium. The polymer stress tensor can be written as, 
\begin{equation}
    \boldsymbol{\tau} = -\eta_p \mathbf{A}_{(1)}, 
    \label{eq:PolymerStressTensor1}
\end{equation}
where $\mathbf{A}_{(1)}$ is the upper convected derivative of the polymer conformation tensor, $\mathbf{A}$. The upper convected derivative of $\mathbf{A}$ is by definition, 
\begin{equation}
 \mathbf{A}_{(1)} = \frac{D}{Dt} \mathbf{A} - \left(\boldsymbol{\nabla} \mathbf{v}\right)^{T} \cdot \mathbf{A} - \mathbf{A}\cdot \left(\boldsymbol{\nabla} \mathbf{v}\right),
 \label{eq:UpperConvected}
\end{equation}
and describes the rate of change of the polymer chain conformation tensor seen by an observer advecting and also rotating and deforming with the flow \cite{rallison1988we}. After incorporating an equilibrium pre-averaging closure approximation, the FENE-P constitutive relation can be written in terms of the dimensionless dumbbell conformation tensor in the form,
\begin{equation}
    \mathbf{A}_{(1)} = -\frac{1}{\lambda}\left[f\left(\textrm{tr}(\mathbf{A})\right)\mathbf{A} - \mathbf{I}\right],
    \label{eq:EFNE-PClosure}
\end{equation}
where
\begin{equation}
    f\left(\textrm{tr}(\mathbf{A})\right) = \frac{L^2 - 3}{L^2 - \textrm{tr}(\mathbf{A})},
    \label{eq:f(trA)}
\end{equation}
$L = r_{max}/\langle r_0^2 \rangle^{1/2}$ is the finite extensibility, $r_{max}$ is the length of a dumbbell at its fully extended state, $\mathbf{I}$ is the identity matrix, and $\lambda$ is the relaxation time of the polymer chain, such that the dimensionless Weissenberg number can be written as $\textrm{Wi} = \lambda \dot{\gamma}$. Combining the FENE-P constitutive equation, \textit{i.e.}, Eq.\,(\ref{eq:EFNE-PClosure}) and (\ref{eq:f(trA)}), with Eq.\,(\ref{eq:PolymerStressTensor1}), the polymer stress tensor for FENE-P fluids can also be written as, 
\begin{equation}
    \tau_{ij} = -\eta_p A_{(1)_{ij}} = G\left[\frac{L^2 - 3}{L^2 - \textrm{tr}(\mathbf{A})}A_{ij} - \delta_{ij}\right], 
    \label{eq:PolymerStressTensor}
\end{equation}
where $G = \eta_p(0)/\lambda$ is the elastic modulus, $\textrm{tr}(\mathbf{A}) = \sum_i A_{ii}$, and $\delta_{ij}$ is the Kronecker delta. It should be noted that there are several variants of the FENE-P model available, as reviewed in \cite{alves2021numerical}.  Similar to \cite{davoodi2022similarities}, we choose the FENE-P variant given above (Eq.(\ref{eq:EFNE-PClosure}) and (\ref{eq:f(trA)})) as the polymer chain conformation tensor reduces to the identity tensor at equilibrium and there is a one-to-one mapping between this variant of the FENE-P constitutive model and the simplified linear Phan-Thien–Tanner (sPTT) model in steady shear flow \cite{davoodi2022similarities}.

We are interested in understanding the evolution of the polymer conformation tensor ($\mathbf{A}$) and polymer stress tensor ($\boldsymbol{\tau}$) with the change in two dimensionless parameters: (i) the Weissenberg number
and (ii) the finite extensibility, $L$. 
To this end, we non-dimensionalize the polymer stress tensor with the elastic modulus such that $\hat{\boldsymbol{\tau}} = \boldsymbol{\tau}/G$.

\section{Analytical, numerical, and asymptotic solutions}
\label{sec:Sol}
For steady simple shear flow, symmetry arguments allow the polymer conformation tensor to be simplified to, 
\begin{equation}
    \mathbf{A} = 
 \begin{pmatrix}
A_{11} & A_{12} & 0\\
A_{21} & A_{22} & 0\\
0 & 0 & A_{33}
\end{pmatrix}.
\end{equation}
The upper convected derivative defined in Eq.\,(\ref{eq:UpperConvected})
can also be simplified to,
\begin{equation}
    \mathbf{A}_{(1)} = \frac{D}{Dt} \mathbf{A} - \dot{\gamma}\begin{pmatrix}
2A_{12} & A_{22} & A_{23}\\
A_{22} & 0 & 0\\
A_{23} & 0 & 0
\end{pmatrix} = \begin{pmatrix}
-2\dot{\gamma}A_{12} & -\dot{\gamma}A_{22} & 0\\
-\dot{\gamma}A_{22} & 0 & 0\\
0 & 0 & 0
\end{pmatrix}.
\end{equation}
Substituting this into the FENE-P constitutive equation (Eq.\,(\ref{eq:EFNE-PClosure}))
provides a set of four nonlinear coupled equations. Solving this set of equations provides the components of the polymer conformation tensor which can then be used to calculate the polymer stress tensor: 
\begin{equation}
    A_{12} = \frac{1}{2\textrm{Wi}}\left[\left(\frac{L^2-3}{L^2 - \textrm{tr}(\mathbf{A})}\right)A_{11}-1\right],
    \label{eq:Nonlinearset1}
\end{equation}
\begin{equation}
    A_{12} = \textrm{Wi}\left(\frac{L^2-\textrm{tr}(\mathbf{A})}{L^2-3}\right)A_{22},
    \label{eq:Nonlinearset2}
\end{equation}
\begin{equation}
    A_{33} = A_{22} = \left(\frac{L^2-\textrm{tr}(\mathbf{A})}{L^2-3}\right),
    \label{eq:Nonlinearset3}
\end{equation}
where $\textrm{tr}(\mathbf{A}) = A_{11}+A_{22}+A_{33} = A_{11} + 2A_{22}$. This set of nonlinear coupled equations can be solved analytically or numerically. An analytical solution has been reported as follows \cite{sureshkumar1997direct},
\begin{equation}
    A_{11} = \frac{1}{F}\left(1+\frac{2\textrm{Wi}^2}{F^2}\right),
    \label{eq:SureshKumar1}
\end{equation}
\begin{equation}
    A_{22}=A_{33} = \frac{1}{F},
    \label{eq:SureshKumar2}
\end{equation}
and
\begin{equation}
    A_{12} = \frac{\textrm{Wi}}{F^2}, 
    \label{eq:SureshKumar3}
\end{equation}
where
\begin{equation}
    F = \frac{\sqrt{3}\Omega}{2\sinh \left(\phi/3\right)}, 
    \label{eq:SureshKumarsub1}
\end{equation}
\begin{equation}
    \phi = \sinh^{-1}\left(\frac{3\sqrt{3}}{2}\Omega\right),
    \label{eq:SureshKumarsub2}
\end{equation}
and
\begin{equation}
    \Omega  = \sqrt{2} \frac{\textrm{Wi}}{L}. 
    \label{eq:SureshKumarsub3}
\end{equation}
This analytical solution can also be directly derived from the form of the FENE-P constitutive equation based on the polymeric extra stress tensor instead of the conformation tensor, as shown in equation (13.5-60) of \cite{bird1987dynamicsv2}. The existence of an analytic solution for the FENE-P model is helpful for computing base flows to be used in turbulent flow simulations or stability analyses. However, it can be hard to understand the asymptotic functional form  of the viscoelastic shear thinning or normal stress response and their dependence on the model parameters from this set of equations. To calculate the asymptotic responses of the polymer conformation and the polymer stress tensor at low and high Weissenberg numbers, we first outline a convenient method to solve Eqs.\,(\ref{eq:Nonlinearset1})-(\ref{eq:Nonlinearset3}) numerically and then evaluate their asymptotic response. We suggest a variable change such that $T = \textrm{tr}(\mathbf{A}) = A_{11}+A_{22}+A_{33}$ and $D = A_{11} - A_{22}$. Thus, $A_{11} = (T+2D)/3$ and $A_{22} = A_{33} = (T-D)/3$. Substituting the values of $A_{11}$, $A_{22}$, and $A_{33}$ into Eqs.\,(\ref{eq:Nonlinearset1})-(\ref{eq:Nonlinearset3}), and solving for $D$, $\textrm{Wi}$, and $A_{12}$ results in the following expressions: 
\begin{equation}
    D = T - 3\left(\frac{L^2-T}{L^2-3}\right) = \frac{L^2\left(T-3\right)}{L^2-3},
    \label{eq:Numerical1}
\end{equation}
\begin{equation}
    \textrm{Wi} = \frac{1}{\sqrt{2}}\left(\frac{L^2-3}{L^2-T}\right)\left[\left(\frac{L^2-3}{L^2-T}\right)\left(\frac{T+2D}{3}\right)-1\right]^{1/2},
    \label{eq:Numerical2}
\end{equation}
and
\begin{equation}
    A_{12} = \textrm{Wi}\left(\frac{L^2-T}{L^2-3}\right)^2.
    \label{eq:Numerical3}
\end{equation}
Equations\,(\ref{eq:Numerical1})-(\ref{eq:Numerical3}) represent three nonlinear coupled equations with four unknowns, \textit{i.e.}, $T$, $D$, $A_{12}$, and $\textrm{Wi}$. A simple way of evaluating them is to recognize that from the physics of the  FENE-P model, the trace of the dimensionless polymer conformation tensor spans over the range $ 3 \leq T \leq L^2$. Defining an equally log-spaced vector for all values that $T$ can take and substituting it in Eq.\,(\ref{eq:Numerical1}), we find a vector for $D$. Substituting the vectors for $T$ and $D$ in Eq.\,(\ref{eq:Numerical2}) provides a vector for the corresponding range of Weissenberg numbers. Finally, substituting the vectors for $T$ and $\textrm{Wi}$ in Eq.\,(\ref{eq:Numerical3}) provides the corresponding values for $A_{12}$. Plots of $T$ vs. Wi, $A_{12}$ vs. Wi, and $D$ vs. Wi can readily be made. From these solutions, we can also readily find the components of the polymer stress tensor. The dimensionless polymer shear stress $\hat{\tau}_{12}$ can be found from Eq.\,(\ref{eq:PolymerStressTensor}) by substituting for $A_{12}$ based on Eq.\,(\ref{eq:Numerical3}) as, 
\begin{equation}
    \hat{\tau}_{12} = \textrm{Wi}\left(\frac{L^2-T}{L^2-3}\right).
    \label{eq:PolyemrShearStressWiLT}
\end{equation}
Using Eq.\,(\ref{eq:PolyemrShearStressWiLT}), the dimensionless polymer viscosity $\eta_{p}(\dot{\gamma})/\eta_{p}(0)$, where $\eta_{p}(0) = G\lambda$, can be evaluated from the dimensionless polymer shear stress and Weissenberg number as, 
\begin{equation}
    \frac{\eta_{P}(\dot{\gamma})}{\eta_{p}(0)} = \frac{\tau_{12}}{\dot{\gamma}}\frac{1}{G \lambda}= \frac{\hat{\tau}_{12}}{\textrm{Wi}}. 
\end{equation}

Using  Eq.\,(\ref{eq:Numerical1}), the dimensionless first normal stress difference $\hat{N}_1$ can be written as,
\begin{equation}
    \hat{N}_1 = \left(\hat{\tau}_{11} - \hat{\tau}_{22}\right) = \left(\frac{L^2-3}{L^2-T}\right)\left(A_{11}-A_{22}\right) = \frac{L^2\left(T-3\right)}{L^2-T}.
    \label{eq:N12}
\end{equation}
Dividing $\hat{N}_1$ by $\textrm{Wi}^2$ gives the dimensionless first normal stress coefficient $\hat{\Psi}_1 = \Psi_1/(G \lambda^2)$, where $\Psi_1$ is the dimensional first normal stress coefficient,
\begin{equation}
    \hat{\Psi}_1 = \frac{\hat{N}_1}{\textrm{Wi}^2} = \frac{\tau_{11} - \tau_{22}}{\dot{\gamma}^2}\frac{1}{G\lambda^2}. 
\end{equation}

Given these analytical expressions and numerical solutions, we can also determine the asymptotic limits connecting the trace of the polymer conformation tensor and different components of polymer stress tensor. We know that $T = 3$ at $\textrm{Wi} = 0$ and consider the ansatz that for a small perturbation from equilibrium, the trace of the conformation tensor has the form,
\begin{equation}
    T \simeq 3+C_1\textrm{Wi}^m, \textrm{ for } \textrm{Wi} \ll 1.
    \label{LowWiAnsatz}
\end{equation}
Combining Eq.\,(\ref{eq:Nonlinearset1}) and (\ref{eq:Nonlinearset3}), we can write, 
\begin{equation}
    2\textrm{Wi}A_{12} = \frac{L^2\left(T-3\right)}{L^2-T}.
    \label{eq:LowWi1}
\end{equation}
Substituting for $A_{12}$ from Eq.\,(\ref{eq:Numerical3}), considering that $L^2 \gg 3$, and keeping terms to the leading order, the low Weissenberg number asymptote for $T$, $D$, and $A_{12}$ are,
\begin{subequations}
\begin{equation}
     \lim_{\textrm{Wi} \ll 1} T = 3+2\textrm{Wi}^2 + O(\textrm{Wi}^4) + \cdots,
    \label{eq:LowWiT}
\end{equation}
\begin{equation}
    \lim_{\textrm{Wi} \ll 1} D = 2\textrm{Wi}^2 + O(\textrm{Wi}^4)+ \cdots,
    \label{eq:LowWiD}
\end{equation}
\begin{equation}
    \lim_{\textrm{Wi} \ll 1} A_{12} = \textrm{Wi} + O(\textrm{Wi}^3)+ \cdots,
    \label{eq:LowWiA12}
\end{equation}
\label{eq:LowWi}
\end{subequations}
respectively. From these results, the low Weissenberg number asymptotes for $\eta_{p}(\dot{\gamma})/\eta_{p}(0)$ and $\hat{\Psi}_1$ are, 
\begin{subequations}
\begin{equation}
     \lim_{\textrm{Wi} \ll 1} \frac{\eta_{P}(\dot{\gamma})}{\eta_{p}(0)} = \frac{\hat{\tau}_{12}}{\textrm{Wi}} = 1 -\frac{2}{L^2-3}\textrm{Wi}^2 + O(\textrm{Wi}^4)+ \cdots,
    \label{eq:LowWiVisc}
\end{equation}
\begin{equation}
     \lim_{\textrm{Wi} \ll 1} \hat{\Psi}_1 = \frac{\hat{N}_{1}}{\textrm{Wi}^2} = 2 - \frac{8}{L^2}\textrm{Wi}^2 + O(\textrm{Wi}^4)+ \cdots,
    \label{eq:LowWiPsi1}
\end{equation}
\label{eq:LowWiOthers}
\end{subequations}
respectively. The low Weissenberg number asymptotes are independent of finite extensibility to the leading order making them universal and equivalent to the results of the Oldroyd-B model.


As $\textrm{Wi} \rightarrow \infty$, $ T \rightarrow L^2$. To find the asymptotic limit of the trace of the conformation tensor in the high Weissenberg number limit, we consider an ansatz of the form, 
\begin{equation}
    T \simeq L^2 - \frac{C_2\left(L^2-3\right)}{\textrm{Wi}^n}.
    \label{eq:HighWiAnsatz}
\end{equation}
The advantage of considering this ansatz is that one can write
\begin{equation}
   \frac{\left(T-3\right)}{\left(L^2-3\right)} \simeq 1-\frac{C_2}{\textrm{Wi}^n}, 
   \label{eq:HighWiAnsatzsub2}
\end{equation}
and 
\begin{equation}
    \frac{\left(L^2-T\right)}{\left(L^2-3\right)} \simeq \frac{C_2}{\textrm{Wi}^n}.
    \label{eq:HighWiAnsatzsub3}
\end{equation}
To find the values of $C_2$ and $n$, we multiply Eq.\,(\ref{eq:Nonlinearset1}) and (\ref{eq:Nonlinearset2}) to eliminate $\textrm{Wi}$ and write $A_{12}$ solely based on $L$, $T$ and other components of the conformation tensor, 
\begin{equation}
    2A_{12}^2 = A_{22}\left(\frac{L^2-T}{L^2-3}\right)\left[\left(\frac{L^2-3}{L^2-T}\right)A_{11}-1\right].
    \label{eq:HighWi1}
\end{equation}
To eliminate $A_{11}$ and $A_{22}$ from Eq.\,(\ref{eq:HighWi1}) and write it solely based on $L$ and $T$, we use a combination of  Eq.\,(\ref{eq:Nonlinearset3}), which provides an expression for $A_{22}$ based on $L$ and $T$, and Eq.\,(\ref{eq:Numerical1}), which provides an expression for $D = (A_{11}-A_{22})$ based on $L$ and $T$. Substituting for $A_{11}$ and $A_{22}$ based on Eq.\,(\ref{eq:Nonlinearset3}) and (\ref{eq:Numerical1}) in Eq.\,(\ref{eq:HighWi1}) gives,
\begin{equation}
    2A_{12}^2 = L^2\left(\frac{T-3}{L^2-3}\right)\left(\frac{L^2-T}{L^2-3}\right).
    \label{eq:HighWi2}
\end{equation}
Substituting the high Weissenberg number asymptote ansatz shown in Eq.\,(\ref{eq:HighWiAnsatz}) in Eq.\,(\ref{eq:HighWi2}) gives, 
\begin{equation}
    2A_{12}^2 \simeq L^2\left(1-\frac{C_2}{\textrm{Wi}^n}\right)\left(\frac{C_2}{Wi^n}\right). 
    \label{eq:HighWi3}
\end{equation}
To leading order, Eq.\,(\ref{eq:HighWi3}) gives an expression for $A_{12}$ of the form,
 \begin{equation}
     A_{12} \simeq \left(\frac{C_2L^2}{2\textrm{Wi}^n}\right)^{1/2}.
     \label{eq:HighWi4}
 \end{equation}
Substituting for $A_{12}$ in Eq.\,(\ref{eq:LowWi1}) using Eq.\,(\ref{eq:HighWi4}) and using the ansatz shown in Eq.\,(\ref{eq:HighWiAnsatz}) for $T$ and simplifying it to leading order gives, 
\begin{equation}
    2\textrm{Wi}\left(\frac{C_2L^2}{2\textrm{Wi}^n}\right)^{1/2} = \frac{L^2}{C_2/\textrm{Wi}^n}. 
\end{equation}
Matching terms requires $1-n/2 = n$  and $\sqrt{C_2L^2/2} = L^2/C_2$ leading to  $n = 2/3$ and $C_2 = \left(L/\sqrt{2}\right)^{2/3}$. Hence the high Weissenberg number asymptotes of $T$, $D$, and $A_{12}$ are:
\begin{subequations}
\begin{equation}
     \lim_{\textrm{Wi} \gg 1} T = L^2 - \left(L^2-3\right)\left(\frac{L}{\sqrt{2} \textrm{Wi}}\right)^{2/3} + O(\textrm{Wi}^{-4/3})+ \cdots,
    \label{eq:HighWiT}
\end{equation}
\begin{equation}
     \lim_{\textrm{Wi} \gg 1} D = L^2 - \left(\frac{L^4}{\sqrt{2}\textrm{Wi}}\right)^{2/3} + O(\textrm{Wi}^{-4/3})+ \cdots,
    \label{eq:HighWiD}
\end{equation}
\begin{equation}
     \lim_{\textrm{Wi} \gg 1} A_{12} = \left(\frac{L^4}{4\textrm{Wi}}\right)^{1/3} + O(\textrm{Wi}^{-1})+ \cdots,
    \label{eq:HighWiA12}
\end{equation}
\label{eq:HighWi}
\end{subequations}
respectively.

\begin{figure}[tbhp]
    \centering
    \includegraphics[width=\columnwidth]{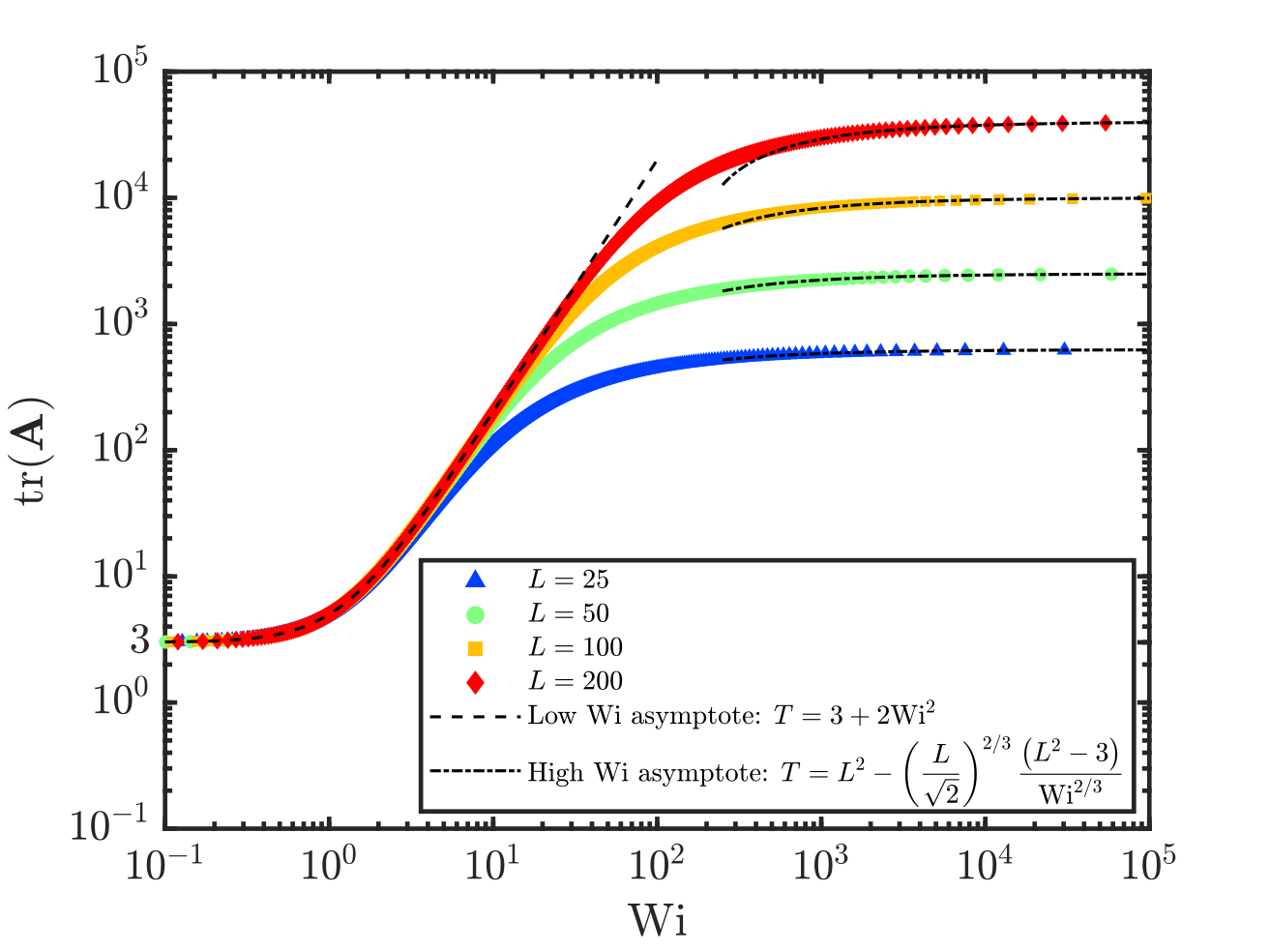}
    \caption{Evolution of the trace of the conformation tensor $\textrm{\textbf{A}}$ with Weissenberg number in steady shear flow for different values of finite extensibility, $L$. The dashed and the dash-dotted lines show the low and high Weissenberg number asymptotes of $\textrm{tr(\textbf{A})}$, respectively.}
    \label{fig:Trace}
\end{figure}

Figure\,(\ref{fig:Trace}) shows the evolution of the trace of the polymer conformation tensor as the Weissenberg number increases from small to large values and for different values of the finite extensibility parameter, $L$. The dashed and the dash-dotted lines show the low and high Weissenberg number asymptotes, respectively. The curves superpose at low Weissenberg number and only start to deviate from each other based on their corresponding extensibility parameter, $L$, when $\textrm{Wi}\gg 1$. The dependence of the deviation from the Oldroyd-B limit on $L$ can also be understood by considering Eq.\,(\ref{eq:LowWi}) more carefully. The higher order term in Equation\,(\ref{eq:LowWiT}) is not just a function of Wi but a function of both Wi and $L$. A more formal way of proceeding would be to recognize that we seek a perturbation away from the known Oldroyd-B solution ($L \rightarrow \infty$) so that $T = 3+2\textrm{Wi}^2( 1  + C_3 (\textrm{Wi}/L)^p + \cdots)$. After some algebra one finds $C_3 = -6$ and $p = 2$, showing that the higher order term represented as $O(\textrm{Wi}^4)$ in Eq.\,(\ref{eq:LowWiT}) is in fact more precisely $-12\textrm{Wi}^4/L^2$. Thus, this term is smaller for FENE-P fluids with higher $L$ resulting in their deviation from the Oldroyd-B result at higher values of Weissenberg number, as shown in Fig.\,\ref{fig:Trace}.

Considering the asymptote of $T$ at high Weissenberg numbers, replacing for $T$ in Eq.\,(\ref{eq:PolyemrShearStressWiLT}) and Eq.\,(\ref{eq:N12}) based on  Eq.\,(\ref{eq:HighWiT}), and keeping the terms to leading order, the high Weissenberg number asymptotes for $\eta_{p}/\eta_{p}(0)$ and $\hat{\Psi}_1$ are, 
\begin{subequations}
\begin{equation}
     \lim_{\textrm{Wi} \gg 1} \frac{\eta_{P}(\dot{\gamma})}{\eta_{p}(0)} = \frac{\hat{\tau}_{12}}{\textrm{Wi}} = 2^{-1/3}\left(\frac{\textrm{Wi}}{L}\right)^{-2/3} + O(\textrm{Wi}^{-4/3})+ \cdots,
    \label{eq:HighWiVisc}
\end{equation}
\begin{equation}
     \lim_{\textrm{Wi} \gg 1} \hat{\Psi}_1 = \frac{\hat{N}_{1}}{\textrm{Wi}^2} = 2^{1/3}\left(\frac{\textrm{Wi}}{L}\right)^{-4/3} + O(\textrm{Wi}^{-2})+ \cdots,
    \label{eq:HighWiPsi1}
\end{equation}
\label{eq:HighWiOthers}
\end{subequations}
respectively.

\begin{figure*}[ht!]
    \centering
    \includegraphics[width=\linewidth]{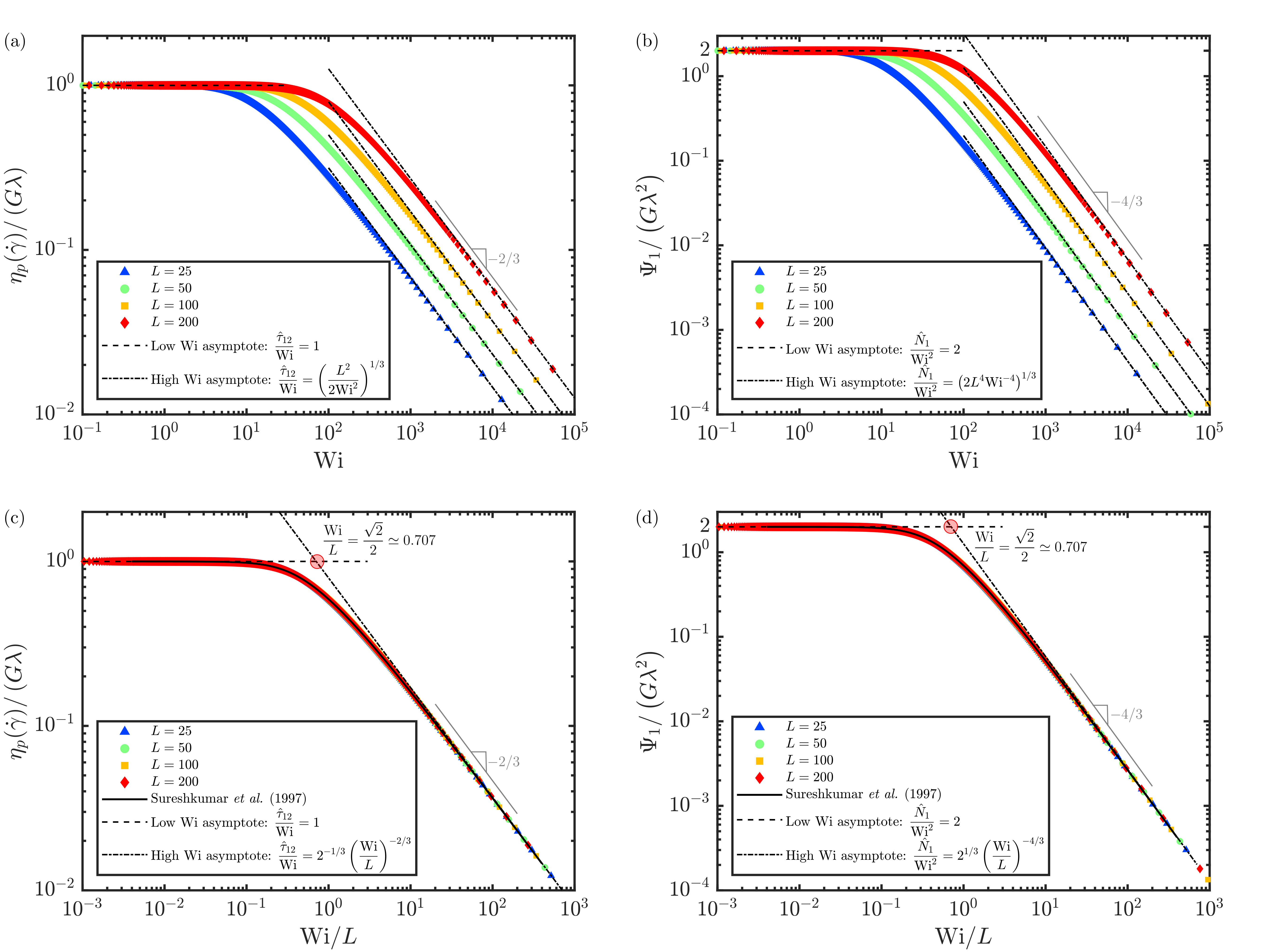}
    \caption{Evolution of (a) the dimensionless polymer viscosity, $\eta_p(\dot{\gamma})/\eta_p(0)$, and (b) the dimensionless first normal stress coefficient $\hat{\Psi}_1$ as the Weissenberg number increases for different values of finite extensibility, $L$. The dashed and the dash-dotted lines show the low and high Weissenberg number asymptotes, respectively. (c) and (d) Master curves collapsing the data shown in (a) and (b), respectively. The solid line shows the analytical solution of Sureshkumar \textit{et al.} \cite{sureshkumar1997direct}, which is faithfully reproduced by our numerical solution. The intersection of the low and high Weissenberg asymptotes (intersection of dashed and dash-dotted lines) at $\textrm{Wi}/L \simeq 0.707$ is marked with a red circle.}
    \label{fig:EtaPsi1}
\end{figure*}

Figures\,\ref{fig:EtaPsi1}(a) and (b) show the evolution of the dimensionless viscosity and the first normal stress coefficient as the Weissenberg number increases, and for different values of the finite extensibility parameter, $L$. The dashed and the dash-dotted lines show the low and high Weissenberg number asymptotes, respectively. Rescaling the abscissa of Fig.\,\ref{fig:EtaPsi1}(a) and (b) by the finite extensibility, $L$, collapses the curves shown in Fig.\,\ref{fig:EtaPsi1}(a) and (b) for different values of $L$ providing master curves for the polymer viscosity and first normal stress coefficient of FENE-P fluids in steady shear flow, as shown in Figs.\,\ref{fig:EtaPsi1}(c) and (d). The ordinate of Fig.\,\ref{fig:EtaPsi1}(c) and (d), however, does not require rescaling as the maximum values of the dimensionless viscosity and the first normal stress coefficient are constants and not a function of Weissenberg number or finite extensibility. The results of our numerical approach are identical to the results of the  analytical solution\,\cite{sureshkumar1997direct}. The low and high Weissenberg number asymptotes, \textit{i.e.}, dashed and dash-dotted lines, respectively, are also accurate approximations in these limits. The master curves presented in Fig.\,\ref{fig:EtaPsi1}(c) and (d) show that the governing dimensionless number in shear flow is not the Weissenberg number \textit{per se} but the ratio of the Weissenberg number to the  finite extensibility parameter, \textit{i.e.}, $\textrm{Wi}/L$. Figures \,\ref{fig:EtaPsi1}(c) and (d) also show that the low and high Weissenberg number asymptotes intersect at $\textrm{Wi}/L \sim 1$, marking the beginning of the region where finite extensibility plays a significant role. At ``high'' Weissenberg numbers, Brownian dynamics simulations using a bead-spring model with FENE springs \cite{lyulin1999brownian,hsieh2004modeling,jendrejack2002stochastic,schroeder2005characteristic} have reported polymer viscosity and first normal stress coefficient following power-law decays with slopes $-2/3$ and $-4/3$, respectively. The FENE-P results are consistent with the Brownian dynamics simulations and the master curves identified for the FENE-P asymptotics show these power-law decays are consistent with a region where $\textrm{Wi}/L \gg 1$, hence providing a more precise definition for ``high'' Weissenberg number. At ``moderate'' Weissenberg numbers, the Brownian dynamics simulations have also reported a less steep power-law decay with slope $-1/2$ and $-14/11$ for polymer viscosity and first normal stress coefficient, respectively \cite{lyulin1999brownian,hsieh2004modeling,jendrejack2002stochastic,schroeder2005characteristic}. These weaker power-law decays can be viewed as locally-valid approximations to the master curve when $\textrm{Wi}/L \sim 1$, hence providing a clearer specification of ``moderate'' Weissenberg number.

Following the same approach, low and high Weissenberg number asymptotes and  master curves can be obtained for $(A_{11} - A_{22})$, $A_{12}$, $(\hat{\tau}_{11} - \hat{\tau}_{22})$, $\hat{\tau}_{12}$, and the stress ratio $(\hat{\tau}_{11} - \hat{\tau}_{22})/\hat{\tau}_{12}$. Similar to Fig.\,\ref{fig:EtaPsi1}(c) and (d), these master curves also have $\textrm{Wi}/L$ for their abscissa. Unlike  Fig.\,\ref{fig:EtaPsi1}(c) and (d), where the maximum values are approached as $\textrm{Wi} \rightarrow 0$, $(A_{11} - A_{22})$ asymptotically approaches a maximum of $L^2$ as $\textrm{Wi}\rightarrow \infty$. In addition, $A_{12}$ passes through a maximum of $L/(2\sqrt{2}) \sim L$ at $Wi = \sqrt{2}L$. Thus the master curves for $(A_{11} - A_{22})$ and $A_{12}$ need their ordinates to be rescaled with $L^2$ and $L$, respectively. The dimensionless first normal stress difference and shear stress are $(\hat{\tau}_{11} - \hat{\tau}_{22}) = f(\textrm{tr}(\mathbf{A}))(A_{11} - A_{22})$ and $\hat{\tau}_{12}= f(\textrm{tr}(\mathbf{A}))A_{12}$, respectively. 
The ordinate of the master curves for the first normal stress difference and shear stress need to be rescaled similar to the ordinate of the master curves for $(A_{11} - A_{22})$ and $A_{12}$, \textit{i.e.}, with $L^2$ and $L$, respectively. Subsequently, the master curve for the stress ratio  $(\hat{\tau}_{11} - \hat{\tau}_{22})/\hat{\tau}_{12}$ has its ordinate rescaled with $L$ and has low and high Weissenberg number asymptotes of   
\begin{subequations}
\begin{equation}
    \lim_{\textrm{Wi} \ll 1} \frac{\hat{\tau}_{11} - \hat{\tau}_{22}}{\hat{\tau}_{12}L} = \frac{2\textrm{Wi}}{L} + O(\textrm{Wi}^3)+ \cdots,
    \label{eq:LowWiStressRatio}
\end{equation}
\begin{equation}
    \lim_{\textrm{Wi} \gg 1} \frac{\hat{\tau}_{11} - \hat{\tau}_{22}}{\hat{\tau}_{12}L} = \left(\frac{4\textrm{Wi}}{L}\right)^{1/3} + O(\textrm{Wi}^{-1/3})+ \cdots,
    \label{eq:HighWiStressRatio}
\end{equation}
\label{eq:StressRatioAsympt}
\end{subequations}
\noindent respectively. While it is not trivial to identify $\textrm{Wi}/L$ as the governing dimensionless number in the  analytical solution \cite{sureshkumar1997direct}, Eq.\,(\ref{eq:SureshKumarsub3}) shows that the ratio $\textrm{Wi}/L$ plays an important role in understanding the simple steady shear flow of FENE-P fluids.

As reviewed in detail by \cite{alves2021numerical}, the FENE-P model is not the only constitutive equation used for modeling dilute polymer solutions. Previous studies using the Phan-Thien–Tanner (PTT) model in channel and pipe flows have shown that  $\epsilon^{0.5} \textrm{De}$ is the important characteristic dimensionless parameter, where De is equivalent of Wi in our work and $\epsilon$ is the (small) parameter regulaizing the response in uniaxial elongation \cite{oliveira1999analytical,cruz2005analytical}. Hence, for the specific case of $\epsilon = 1/L^2$, $\epsilon^{0.5} \textrm{De} \equiv \textrm{Wi}/L$. It was recently shown that by setting $\epsilon = 1/L^2$, the sPTT model in steady shear flow provides identical results to the  FENE-P model \cite{davoodi2022similarities}. Thus, the master curves shown in this work can also be extended to the sPTT model for the specific case of $\epsilon = 1/L^2$.

\section{Experimental results}
\label{sec:Exps}

\begin{table*}
  \caption{Chemical composition and rheological properties of the dilute solution of polystyrene in oligomeric styrene.}
   \label{tab:PS}
   \centering
   \resizebox{0.99\textwidth}{!}{
  \begin{tabular}{ccccccccccc}
\hline\hline
\\[-1em]
Sample ID & Polystyrene (ppm) & $Mw$ (g/mol) & \multicolumn{1}{c}{PDI} & \multicolumn{1}{c}{$R_g \textrm{ (nm)}$} & \multicolumn{1}{c}{$L$} & $\lambda \textrm{ (ms)}$ & $\eta_0 \textrm{ (Pa}\cdot\textrm{s)}$ & $\eta_s \textrm{ (Pa}\cdot\textrm{s)}$ & $\Psi_{10} \textrm{ (Pa}\cdot\textrm{s}^2)$ & $G \textrm{ (Pa)}$ \\ 
\hline
\\[-1em]
PS  & $500$        & $2\times 10^6$                            & $1.03$  & $41$                                              & $21$    & $3190$ & $39.2$ & $34$ & $20.3$ & $1.63$                                          \\
\hline\hline
  \end{tabular}
  }
\end{table*}

In this section, we compare the master curves calculated in Sec.\,\ref{sec:Sol} to experimental measurements in steady shear flow using three dilute polymer solutions with different finite extensibility parameters. We first consider the experimental results shown in \cite{anna2001interlaboratory}. The viscosity and the first normal stress coefficient of a dilute solution of narrow-molecular-weight-distribution polystyrene in oligomeric stryrene were measured in steady shear flow over several decades of shear rate and at several temperatures. Table \ref{tab:PS} shows the composition and rheological properties of this solution that we refer to as PS. The narrow molecular weight distribution of the polystyrene used in PS results in a low polydispersity index (PDI) of 1.03. Hence, a well-defined, single radius of gyration can be written for the polymer molecules of this solution. Knowing $R_g$ and the length of the polymer chain at its fully extended state and given $\langle r_0^2 \rangle^{1/2} = \sqrt{6}R_g$ \cite{larson2005rheology}, a single finite extensibility parameter $L$ can be calculated for PS. The longest  relaxation time $\lambda$ of the polymer molecules in this solution was measured using a capillary breakup extensional rheometer (CaBER) \cite{Anna2001}, and the zero shear rate viscosity $\eta_0$ and first normal stress coefficient $\Psi_{10}$ were calculated by fitting the Zimm model \cite{bird1987dynamicsv1, larson2013constitutive} to the experimentally measured dynamic moduli of the dilute polymer solution \cite{anna2001interlaboratory}. It should be noted that the difference between the finite extensibility parameter reported for PS in our work and  in \cite{anna2001interlaboratory} is due to the subtle but important numerical differences in the definition of this parameter. Specifically, the finite extensibility parameter in \cite{anna2001interlaboratory} was defined as $L_{\textrm{Anna}} = (\sqrt{3}r_{max})/R_g$. Using the result $R_g  = \langle r_0^2 \rangle^{1/2}/\sqrt{6}$ and the definition below Eq.\,(\ref{eq:f(trA)}), we obtain our definition of the finite extensibility parameter $L = L_{\textrm{Anna}}/(3\sqrt{2})$.

In addition to PS, we have carried out steady shear flow experiments for two dilute polymer solutions with different solvent and solutes compared to PS. Solutions of hydrolyzed polyacrylamide (HPAM) with 30\% carboylated monomers (Poly(acrylamide/sodium acrylate) [70:30], Polysciences) with reported molecular weight $M_w = 18 \times 10^6 \textrm{ g/mol}$ in a mixture of glycerol, water, and dimethyl sulfoxide (DMSO) are studied. The composition of the two solutions are summarized in Table\,\ref{tab:samples}. We refer to these samples as HPAM-1 and HPAM-2, and more details on their preparation and rheological properties can be found in \cite{browne2020bistability} and \cite{browne2021elastic}, respectively. For HPAM, given $71.07 \textrm{ g/mol}$ as the average molecular weight of the repeat unit, the number of repeat units in each polymer chain is $n \simeq 250,000$. Each repeat unit has four C-C bond with length $\ell = 1.54~ \AA$ for each C-C bond. Thus the length of fully extended polymer chain is $r_{max} = 0.82\times 4\times \ell \times n = 128 \textrm{ } \mu\textrm{m}$. Note that the numerical factor is due to the tetrahedral configuration of the C-C bonds at the fully extended state. Dynamic light scattering is used to measure the equilibrium radius of gyration, $R_g$ of the polymer chains. Unlike PS, HPAM is polydisperse resulting in the measured radii of gyration having a broad distribution, \textit{i.e.}, $133 \textrm{ nm}<R_g< 423 \textrm{ nm} $ and $ 57 \textrm{ nm}<R_{g}< 387 \textrm{ nm}$ for  HPAM-1 and HPAM-2, respectively. This polydispersity results in a distribution in values of the finite extensibility parameter, \textit{i.e.},  $123<L<392$  and $135<L<917$ for samples HPAM-1 and HPAM-2, respectively. In this work, we use the mean radius of gyration of each sample to calculate an average finite extensibility parameter. These average values are reported in Table\,\ref{tab:samples}. In addition, the longest relaxation time, $\lambda$, of the samples are directly measured using a capillary breakup extensional rheometer (CaBER) \cite{oliveira2006iterated} and reported in Table\,\ref{tab:samples}. The errors reported in the table for extensional relaxation time are calculated based on multiple CaBER measurements (five tests for each sample). Steady shear flow experiments are carried out on the samples with a 60 mm $1.02^{\circ}$ acrylic cone and plate geometry using a controlled stress rheometer (DHR3, TA Instruments). Shear stress and first normal stress difference are measured for $10 \textrm{ s}^{-1} < \dot{\gamma} < 1000 \textrm{ s}^{-1}$. The choice of a viscous solvent for both samples allows for rheological measurements with cone and plate geometry over two decades of shear rate.    

\begin{table*}
  \caption{Chemical composition and rheological properties of dilute solutions of HPAM in different solvents.}
   \label{tab:samples}
   \centering
   \resizebox{0.99\textwidth}{!}{
  \begin{tabular}{ccccccccccc}
\hline\hline
\\[-1em]
Sample ID & HPAM (ppm) & NaCl (\%wt.) & \multicolumn{1}{c}{Glycerol (\%wt.)} & \multicolumn{1}{c}{Water (\%wt.)} & \multicolumn{1}{c}{DMSO (\%wt.)} & $\bar{R}_g \textrm{ (nm)}$ & $\bar{L}$ & $\lambda \textrm{ (ms)}$ & $\eta_0 \textrm{ (Pa}\cdot\textrm{s)}$ & $G \textrm{ (Pa)}$ \\ 
\hline
\\[-1em]
HPAM-1  & 300        & 1                            & 89.0                                                 & 10.0                                              & 0.0    & $278$ & $188$ & $3461 \pm 139$ & $0.20 \pm 0.01$ & $0.02 \pm 0.01$                                          \\
HPAM-2  & 300        & 1                            & 82.6                                                 & 6.0                                               & 10.4 & $222$ & $235$ & $498 \pm 6$    &  $0.17 \pm 0.01$ &  $0.04 \pm 0.01$                                      \\
\hline\hline
  \end{tabular}
  }
\end{table*}

\begin{figure}
    \centering
    \includegraphics[width=\columnwidth]{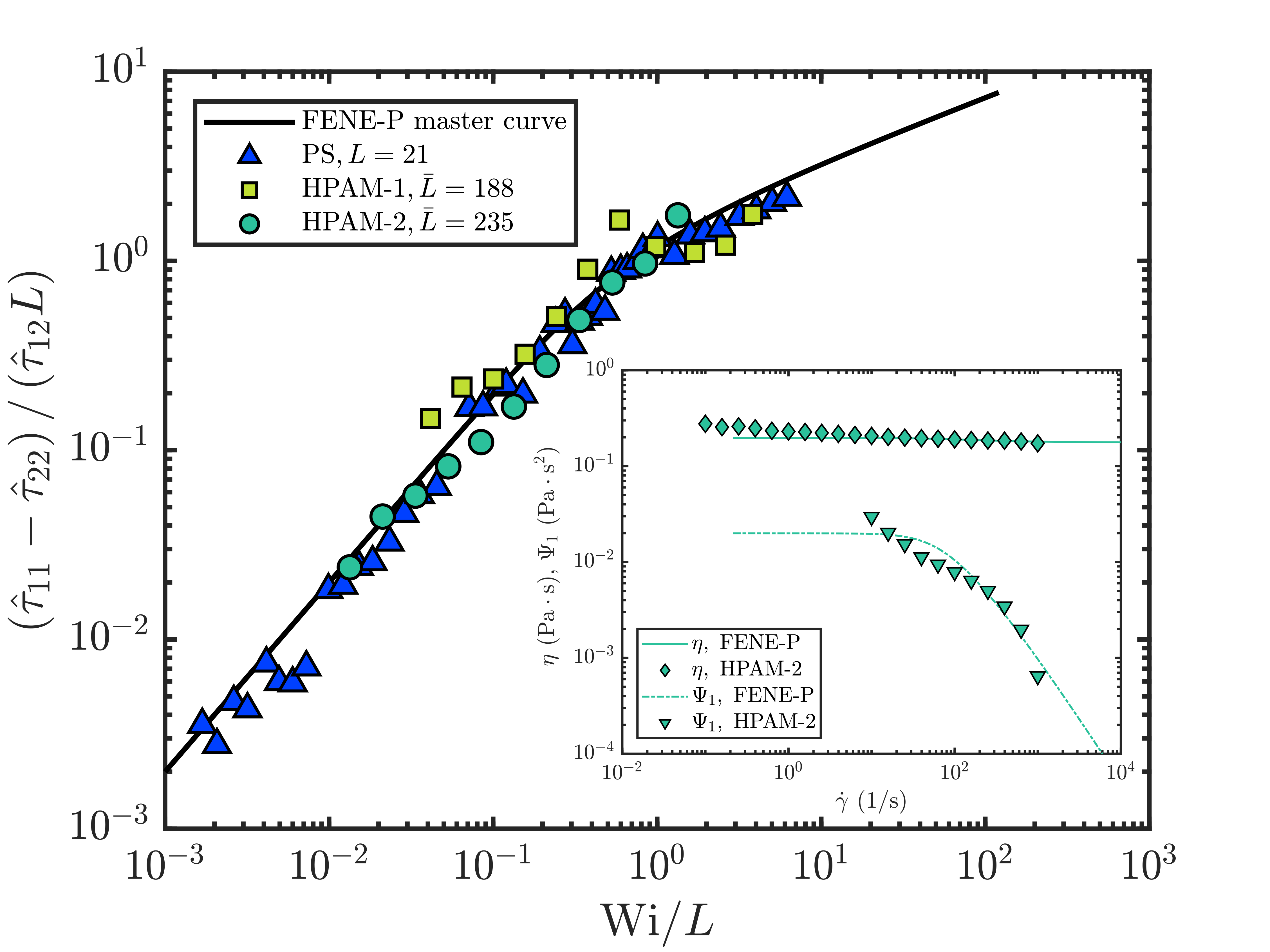}
    \caption{
    Comparison of experimental measurements with the FENE-P master curve for the ratio of the first normal stress difference to the shear stress when scaled with the finite extensibility parameter (shown as solid line). The inset shows the comparison of the dimensional viscosity and the first normal stress coefficient for sample HPAM-2 with the predictions of the FENE-P model based on the fitting parameters shown in Table\ref{tab:samples}.}
    \label{fig:Exps}
\end{figure}

Figure\,\ref{fig:Exps} compares the experimental results with the FENE-P master curve for the ratio of first normal stress difference to the shear stress, shown with solid line. The symbols show this stress ratio for the three solutions and for different values of Weissenberg number, $\textrm{Wi} = \lambda \dot{\gamma}$, where $\lambda$ is the relaxation time measured by CaBER and reported in Tables\,\ref{tab:PS} and \ref{tab:samples}. The normal stress difference measured by the axial force transducer of the rheometer is divided by the polymer contribution to the shear stress, $\tau_{12}$ and by the FENE parameter $L$ to get the ordinate. To find the polymer shear stress for HPAM-1 and HPAM-2, the viscosity of the solvent is multiplied by the shear rate to provide the solvent contribution to the shear stress. This is then subtracted from the total shear stress measured by the rheometer to find the polymer contribution, $\tau_{12}$. In addition to this dimensionless comparison, the inset of Fig.\,\ref{fig:Exps} compares the dimensional properties of one of the HPAM solutions with the predictions of the FENE-P model. The measured value of the zero shear rate viscosity is used as the fitting parameter, together with the relaxation time and the solvent viscosity to redimensionalize the dimensionless FENE-P predictions. The inflection point that is noticeable in the first normal stress coefficient of HPAM-2 is due to additional effects associated with anisotropic drag on the polymer molecules not captured by the FENE model \cite{bird1985anisotropic, magda1988deformation} and has previously been observed for other Boger fluids \cite{quinzani1990modeling}.

The three sets of experimental data, which correspond to three different values of relaxation time and finite extensibility, collapse quite well on the master curve calculated from the FENE-P theory. Despite the simplicity of the FENE-P model, the agreement between experiments and the FENE-P master curve illustrates the model's utility in capturing the macroscopic dynamics of dilute solutions of high molecular weight flexible polymers.

\section{Governing dimensionless number}
\label{sec:MasterCurve}

Our theoretical and experimental results show that the governing dimensionless number for the master curves of FENE-P fluids in steady shear flow is the ratio of the Weissenberg number to the finite extensibility parameter, $\textrm{Wi}/L$. In this section we provide a physical picture illustrating why $\textrm{Wi}/L$ is the governing dimensionless number in shear flows but not in purely extensional flows. While the FENE-P is only an ensemble-averaged spring-dumbbell model, it is able to emulate the key features of the physical picture outlined below.

The key additional characteristic of shear flow is a non-zero vorticity.  The existence of vorticity results in tumbling of the individual polymer chains \cite{lumley1969drag,de1974coil, teixeira2005shear}. Tumbling motion changes the relative locations of the two ends from which the polymer chain is stretched. The polymer chain thus continuously tumbles and then stretches again in the shear field. This is in contrast to purely extensional flows, which are vorticity-free and hence tumbling does not occur. Under an extensional strain, the ends of the of the polymer chain continually separate in time until stretching ceases or a Weissenberg-number-dependent steady state is reached. The lack of internal degrees of freedom in the dumbbell model means it cannot capture polymer chain tumbling. The FENE-P force law, however, mimics the key result by making the first normal stress difference grow as $L^{4/3}$ in a steady shear flow, as shown in Eq.\,(\ref{eq:HighWiPsi1}) and predicted by bead-rod \cite{doyle1997dynamic, hur2001dynamics} and bead-spring \cite{hsieh2004modeling,lyulin1999brownian, schroeder2005dynamics, schroeder2005characteristic} simulations, only when $\textrm{Wi}/L \gg 1$. The ratio $\textrm{Wi}/L$ is thus the governing dimensionless number distinguishing ``weak linear shearing flows'' in which the nonlinear effects of finite chain extensibility are not important and ``strong nonlinear shearing flows'' in which finite extensibility and chain tumbling play a role. In contrast, the steady state extensional stress difference in purely extensional flow grows as $L^2$ for all Weissenberg numbers beyond the coil-stretch transition at $\textrm{Wi} = 1/2$ \cite{bird1987dynamicsv2} making the ratio  $\textrm{Wi}/L$ irrelevant. The slower power-law growth with finite extensibility of the first normal stress difference in shear flow compared to extensional flow is a result of the asymptotic form of the FENE-P equations at $\textrm{Wi}/L \gg 1$, which prevents the dumbbell from reaching a fully stretched state and increasingly aligns it in the flow direction. 

As the Weissenberg number increases, the increasingly strong flow stretches the polymer chains and progressively aligns them with the flow direction, reducing the mean orientation angle. In this flow-aligned state, small Brownian fluctuations can make the local angle with the flow momentarily negative and induce a molecular tumbling event \cite{schroeder2005characteristic}. As Weissenberg number increases, polymer chains thus experience higher stretch (when the orientation angle is positive), with a maximum achievable stretch that depends on $L$. At the same time the frequency of tumbling also increases \cite{teixeira2005shear}. These two effects are in competition, as tumbling changes the relative locations of the two ends from which the polymer chain is stretched and resets the chain stretch. This simple physical picture and a more detailed analysis of the physics of repeated tumbling and stretching show that the parameter $\textrm{Wi}/L$ controls the relative magnitude of the two effects. 

\section{Conclusions}
\label{sec:Conclusion}

We have calculated the polymer contributions to the stresses in steady shear flow and their asymptotic forms at low and high Weissenberg numbers using the FENE-P model. These stresses are dependent on the finite extensibility of the polymer chains at high Weissenberg numbers and exhibit different power-laws compared to purely extensional flows. We identify master curves that collapse the polymer stresses and corresponding material properties such as the viscosity and first normal stress coefficient plots independent of the finite extensibility parameter when plotted against $\textrm{Wi}/L$. Previous direct Navier-Stokes simulations using the FENE-P model in shear-dominated flows have studied the effect of changes in Weissenberg number and the finite extensibility parameter on the flow independently \cite{liu2013polymer, shaqfeh2021oldroyd,page2016viscoelastic,page2015dynamics}. We show that the governing dimensionless number for FENE-P fluids in shear flows is the ratio of the Weissenberg number to the finite extensibility. This ratio of parameters emulates how much stretch a polymer chain can accumulate between its tumbling motions compared with the maximum stretch that a polymer chain possesses due to its finite extensibility. In the limit $\textrm{Wi}/L \ll 1$ the expected Oldroyd-B scalings for $\hat{N}_1$ and $\hat{\tau}_{12}$ are recovered. The finite extensibility parameter, however, plays a significant role in controlling the asymptotic form of the polymer stress tensor when $\textrm{Wi}/L \gtrsim 1$. The FENE-P model has been widely used in recent computations of viscoelastic free and wall-bounded shear flows, especially at moderate to high Reynolds numbers \cite{buza2022weakly,shekar2021tollmien, lopez2019dynamics, guimaraes2020direct, parvar2022steady, lopez2022vortex, guimaraes2022turbulent, dubief2022first} where molecular dynamics simulations are computationally expensive and an ensemble-averaged closure equation is needed to calculate the polymer stress tensor. Based on our results, it would seem to be helpful to report the value of the key governing dimensionless number $\textrm{Wi}/L$ when using the FENE-P model in shear flow computations to clarify whether the shear rate is high enough for the finite extensibility of the polymer chains to have a significant impact on the polymer stress tensor.

\section{Acknowledgements}
 We are grateful to Dr. Bavand Keshavarz for his valuable insight and detailed technical suggestions. We are also grateful to Dr. Christopher A. Browne and Professor Sujit S. Datta for sharing experimental samples (HPAM-1 and HPAM-2) and dynamic light scattering measurements with us. We also acknowledge the insightful comments and suggestions we received from the reviewers during the peer review process. This work was supported by the National Science Foundation (NSF) Grant No. CBET-2027870 to MIT. We also acknowledge the support of the Natural Sciences and Engineering Research Council of Canada (NSERC), funding reference number CGSD2-532512-2019. Cette recherche a été financée par le Conseil de recherches en sciences naturelles et en génie du Canada (CRSNG), numéro de référence CGSD2-532512-2019. 


\bibliographystyle{elsarticle-num.bst}
\bibliography{references.bib}


\end{document}